\documentclass[12pt,twoside]{article}
\usepackage{graphicx,color}
\usepackage{amsmath}
\newcommand\pubnumber{TTP11-03\\TUM-HEP-792/11}
\newcommand\pubdate{February 2011}

\def\napoli{
      Institut f{\"u}r Theoretische Teilchenphysik
      \\
      Karlsruhe Institute of Technology,
      Universit{\"a}t Karlsruhe
      \\
      Engesserstra\ss e 7, 76128 Karlsruhe, Germany
}
\def\munich{
      Physik Department
      \\
      Technische Universit{\"a}t M{\"u}nchen
      \\  
      James-Franck-Str, D-85748 Garching, Germany
}

\def\Title#1{\begin{center} {\Large #1 } \end{center}}
\def\Author#1{\begin{center}{ \sc #1} \end{center}}
\def\Address#1{\begin{center}{ \it #1} \end{center}}
\def\andauth{\begin{center}{and} \end{center}}

\newcommand\pubblock{\rightline{\begin{tabular}{l} \pubnumber\\
         \pubdate  \end{tabular}}}
\newenvironment{Abstract}{\begin{quotation}  }{\end{quotation}}
\newenvironment{Presented}{\begin{quotation} \begin{center} 
             PRESENTED AT\end{center}\bigskip 
      \begin{center}\begin{large}}{\end{large}\end{center} \end{quotation}}
\def\Acknowledgements{\bigskip  \bigskip \begin{center} \begin{large}
             \bf ACKNOWLEDGEMENTS \end{large}\end{center}}
 \newcommand{\ket}[1]{|#1\rangle}

\newcommand{\lt}{\left}
\newcommand{\rt}{\right}

\newcommand{\gev}{\,\mbox{GeV}}
\newcommand{\mev}{\,\mbox{MeV}}

\newcommand{\Bbar}{\,\overline{\!B}}
\newcommand{\bbq}{$\mathrm{B_q}\!-\!\ov{\mathrm{B}}{}_\mathrm{q}\,$}
\newcommand{\bbd}{$\mathrm{B_d}\!-\!\ov{\mathrm{B}}{}_\mathrm{d}\,$}
\newcommand{\bbs}{$\mathrm{B_s}\!-\!\ov{\mathrm{B}}{}_\mathrm{s}\,$}
\newcommand{\bbms}{$\mathrm{B_s}\!-\!\ov{\mathrm{B}}{}_\mathrm{s}\,$\ mixing}
\newcommand{\bbmd}{$\mathrm{B_d}\!-\!\ov{\mathrm{B}}{}_\mathrm{d}\,$\ mixing}
\newcommand{\bbmq}{$\mathrm{B_q}\!-\!\ov{\mathrm{B}}{}_\mathrm{q}\,$\ mixing}

\newcommand{\bb}{$\mathrm{B}\!-\!\ov{\mathrm{B}}{}\,$}
\newcommand{\bbm}{$\mathrm{B}\!-\!\ov{\mathrm{B}}{}\,$\ mixing}
\newcommand{\kkm}{$\mathrm{K}\!-\!\ov{\mathrm{K}}{}\,$\ mixing}

\newcommand{\dm}{\ensuremath{\Delta M}}
\newcommand{\dg}{\ensuremath{\Delta \Gamma}}
\newcommand{\BsorBsbar}{\raisebox{7.7pt}{$\scriptscriptstyle(\hspace*{8.5pt})$}
  \hspace*{-10.7pt}\!\Bbar_{s}}

\newcommand{\BqorBqbar}{\raisebox{7.7pt}{$\scriptscriptstyle(\hspace*{8.5pt})$}
  \hspace*{-10.7pt}\!\Bbar_{q}}

\newcommand{\eq}[1]{Eq.~(\ref{#1})}
\newcommand{\eqsand}[2]{Eqs.~(\ref{#1}) and (\ref{#2})}

\newcommand{\bea}{\begin{eqnarray}}
\newcommand{\eea}{\end{eqnarray}}

\newcommand{\nn}{\nonumber \\}
\newcommand{\no}{\nonumber}
\newcommand{\ov}{\overline}
\newcommand{\epm}[2]{
 \raisebox{-0.5ex}{\shortstack[l]{$\scriptstyle+#1$\\$\scriptstyle-#2$}}}

\def\journal#1#2#3#4{#1~{\bf #2}, #3 (#4)}

\def\PLB#1#2#3{\journal{Phys.\ Lett. B}{#1}{#2}{#3}}

\def\NPB#1#2#3{\journal{Nucl.\ Phys. B}{#1}{#2}{#3}}

\def\PRD#1#2#3{\journal{Phys.\ Rev. D}{#1}{#2}{#3}}
\def\PRL#1#2#3{\journal{Phys.\ Rev. Lett.}{#1}{#2}{#3}}

\newcommand{\arxiv}[1]{{arxiv:{#1}}}
\newcommand{\fig}[1]{Fig.~\ref{#1}}

\begin{document}
\begin{titlepage}
\pubblock

\vfill
\Title{\bf Numerical updates of lifetimes and\\[2mm] 
           mixing parameters of B mesons}
\vfill
\Author{Alexander Lenz}
\Address{\munich}
\andauth
\Author{Ulrich Nierste}
\Address{\napoli}
\vfill
\begin{Abstract}
  {We update the Standard-Model predictions for several quantities
  related to \bbs\ and \bbmd. The mass and width differences in the 
  $B_s$ system read $\dm_s^{\rm SM} = (17.3 \pm 2.6) \,
  \mbox{ps}^{-1}$ and $\dg_s^{\rm SM}\;=\;  
   (0.087 \pm 0.021 )\, \mbox{ps}^{-1}$, respectively. The CP
   asymmetries in flavour-specific decays are  
   $a_{\rm fs}^{s,\rm SM} = (1.9 \pm 0.3) \cdot 10^{-5}$ and 
   $a_{\rm fs}^{d,\rm SM} = - (4.1 \pm 0.6) \cdot 10^{-4}$. We further
   critically discuss the sensitivity of $\dg_d$ to new physics and 
   the uncertainties in the relation between  $\dg_s$ and the branching
   fraction of
   $\BsorBsbar \to D_s^{(*)}{}^+ D_s^{(*)}{}^-$. Then we
   present a numerical update of the average width $\Gamma_s$ in the
   $B_s$ system {and correlate $\Gamma_s$ with the $B^+$--$B_d$
     lifetime ratio. Finally we}
   summarise the key results of our recent global 
   analysis with the CKMfitter collaboration addressing new physics in \bbm. 
   In an appropriately defined scenario parametrising new physics in
   \bbm\ by two complex parameters $\Delta_d$ and $\Delta_s$ the
   Standard-Model point {$\Delta_d=\Delta_s=1$} is disfavoured by 3.6
   standard deviations. }
\end{Abstract}
\vfill
\begin{Presented}
6th International Workshop on the CKM Unitarity Triangle\\  
Warwick, United Kingdom, September 6-10, 2010
\end{Presented}
\vfill
\end{titlepage}
\def\thefootnote{\fnsymbol{footnote}}
\setcounter{footnote}{0}

\section{Introduction}
On May 14, 2010, the D\O\ collaboration has reported evidence for an
anomalous like-sign dimuon charge asymmetry $A_\text{SL}$ in the decays
of neutral $B$ mesons \cite{dimuon_evidence_d0}.  The presented result
corresponds to a data set composed of $B_d$ and $B_s$ mesons and
quantifies CP violation in the \bbd\ and \bbms\ amplitudes.  Expressed
in terms of the CP asymmetries $a_{\rm fs}^{d,s}$ in flavour-specific
$B_{d,s}$ decays the measured quantity reads 
\begin{eqnarray}
A_\text{SL} &=& (0.506 \pm 0.043) a_{\rm fs}^{d} + 
 (0.494 \pm 0.043) a_{\rm fs}^{s} .\label{asl}
\label{DEFdimuon}
\end{eqnarray}
The index $\text{SL}$ refers to the use of semileptonic decays in the
measurement.  The D\O\ result $A^{\text{D\O}}_\text{SL}=-0.00957 \pm
0.00251 \pm 0.00146$ deviates from the Standard-Model (SM) prediction of
Ref.~\cite{ln}, $A_\text{SL}=\lt( -0.23\epm{0.05}{0.06} \rt) \cdot
10^{-3}$, with a statistical significance of 3.2 standard deviations and
the central value is off by a factor of 42.  With the direct and
indirect knowledge on $ a_{\rm fs}^{d}$ from $B$ factory data (which
still leaves room for sizeable new physics contributions) any
theoretical explanation of the D\O\ measurement necessarily requires
large new physics contributions to the \bbms\ amplitude. This
contribution must be similar in magnitude to the SM contribution while
having a very different phase.

\bbmq\ is described by two hermitian $2\times 2$ matrices, the mass
matrix $M^q$ and the decay matrix $\Gamma^q$. By diagonalising $M^q -
i \Gamma^q/2$ one determines the mass eigenstates $\ket{B_q^H}$ and
$\ket{B_q^L}$ (with ``H'' and ``L'' denoting ``heavy'' and ``light'')
as linear combinations of the flavour eigenstates $\ket{B_q}$ and
$\ket{\Bbar_q}$.  The average mass of the two eigenstates is
$M_{B_q}=M^q_{11}=M^q_{22}$ and the average width equals
$\Gamma_q = \Gamma^q_{11}=\Gamma^q_{22}$.
%\begin{eqnarray} 
%\Gamma_q &=& \Gamma^q_{11}=\Gamma^q_{22} .  \label{aw}
%\end{eqnarray}
The off-diagonal elements $M^q_{12}$ and $\Gamma^q_{12}$ lead to 
\bbmq\ phenomena, namely a mass difference $\dm_q$ and a width 
difference $\dg_q$ between the eigenstates $B_q^H$ and $B_q^L$ and
further to the CP asymmetry in flavour-specific decays, $a_{\rm fs}^{q}$. 
These quantities read 
\begin{eqnarray}
\dm_q & = & M_H^q-M_L^q \; \simeq \; 2 |M_{12}^q| \, , \qquad\qquad
\dg_q \;=\; \Gamma_L^q-\Gamma_H^q \;\simeq \; 2 |\Gamma_{12}^q| \cos
\phi_q  \, ,\nn
a_{\rm fs}^{q} &=& 
\frac{|\Gamma_{12}^q|}{|M_{12}^{q}|} \sin \phi_q \, ,\no
\end{eqnarray}
with the CP-violating phase 
\begin{eqnarray}
 \phi_q &\equiv& \arg \lt( - \frac{M_{12}^q}{\Gamma_{12}^q} \rt). \label{defphi}
\end{eqnarray}
For pedagogical introductions to \bbmq\ see Refs.~\cite{ped}.
In the SM the CP phases are small,  
$\phi_d\approx {-4.3^\circ}$ and 
$\phi_s\approx {0.22^\circ}$, so that 
$\dg_q^{\rm SM}\simeq  2 |\Gamma_{12}^q|$. New physics can affect
magnitude and phase of $M_{12}^q$ and $\dm_q$ and $\phi_q$ 
can deviate from their SM predictions substantially. 

We present updates of the Standard-Model predictions for $\dg_q$ and
{the CP asymmetries in flavour-specific decays in Sec.~\ref{sec:dg}. 
Sec~\ref{sec:gs} is devoted to $\Gamma_s$ and}
in Sec.~\ref{sec:np} we discuss the recent analysis of \bbm\ in
Ref.~\cite{lnckmf}, which shows evidence of new physics.  

\boldmath 
\section{Width differences {and CP asymmetries}}\label{sec:dg}
\unboldmath The calculation of $\Gamma^q_{12}$ uses the heavy quark
expansion (HQE), which is an operator product expansion exploiting the
hierarchy $m_b \gg \Lambda_{\rm QCD}$. $\Gamma^q_{12}$ is then predicted
as a simultaneous expansion in the two parameters $\Lambda_{\rm
  QCD}/m_b$ and $\alpha_s(m_b)$. $\BqorBqbar$ decays into final states
which are common to $B_q$ and $\Bbar_q$ contribute to $\Gamma^q_{12}$
and lead to a width difference between the two eigenstates of the \bbq\
complex.  $\Gamma^s_{12}$ is dominated by the Cabibbo-favoured
contribution with a $(c\ov c)$ pair in these final states.  Corrections
of order $\Lambda_{\rm QCD}/m_b$ to $\Gamma^s_{12}$ have been found in
Ref.~\cite{bbd}, the contributions of order $\alpha_s(m_b)$ have been
calculated in Ref.~\cite{bbgln} and were confirmed in
Ref.~\cite{rome}. The theory prediction of $\Gamma^d_{12}$ to NLO in
$\alpha_s$ and $1/m_b$ has been obtained in Refs.~\cite{rome,bbln}.  In
Ref.~\cite{ln} these NLO results for $\Gamma^q_{12}$ have been expressed
in terms of a new operator basis, which avoids certain numerical
cancellations and improves the NLO results by including some
color-enhanced $\alpha_s/m_b$ corrections. Further Ref.~\cite{ln}
applies the all-order summation of $\alpha_s^n z \ln^n z$ terms (with
$z=m_c^2/m_b^2$ and $n=1,2,\ldots$) of Ref~\cite{bbgln2} to
$\Gamma^q_{12}$. 

In the ratio $|\Gamma_{12}^q|/|M_{12}^{\rm SM,q}|$ hadronic
uncertainties cancel to a large extent. In the 
$B_d$ system on has \cite{ln}  
\begin{eqnarray}
\frac{|\Gamma_{12}^d|}{|M_{12}^{d,\rm SM}|} \, =\, 
\frac{2|\Gamma_{12}^d|}{|\dm_d^{\rm SM}|} 
&=& \lt( {54} \pm 10
\rt) \cdot 10^{-4} \label{dgdmd}
\end{eqnarray}
In the absence of new physics we can identify $\dm_d^{\rm SM}$ with
the experimental value $\dm_d^{\rm exp}= 0.507 \, \mbox{ps}^{-1}$ to
find 
\begin{eqnarray}
 \lt. \frac{\dg_d}{\Gamma_d} \rt|_{\rm SM} &=& 
  \lt( {42} \pm 8 \rt) \cdot 10^{-4}. \label{dgdsm}
\end{eqnarray} 
Here also $\tau(B_d)=1/\Gamma_d=(1.525\pm 0.009)\, \mbox{ps}$ has been
used.  {In the presence of new physics \eq{dgdsm} changes to
\begin{eqnarray}
 \lt. \frac{\dg_d}{\Gamma_d} \rt.  &=& 
  \lt( 45 \pm 10 \rt)  \cdot 10^{-4} \cdot
     \cos \phi_d . 
 \label{dgdnp}
\end{eqnarray}
Different central values in \eqsand{dgdsm}{dgdnp} occur, because we
  do not use $\dm_d^{\rm exp}$ in \eq{dgdnp} and $\dm_d$ is about 7\%
  larger than $\dm_d^{\rm exp}$ for our range of the relevant hadronic
  parameter, $f_{B_d}\sqrt{{\cal B}_{B_d}}= (174\pm 13) \,\mev$.}
Using the 3$\sigma\,\rm CL$ range in Tab.~11 of Ref.~\cite{lnckmf},  
$-30^\circ \leq \phi_d \leq -1^\circ$, one finds that 
\begin{eqnarray}
 \lt. \frac{\dg_d}{\Gamma_d} \rt.  &=& 
  \lt( 45 \epm{10}{12} \rt)  \cdot 10^{-4}
  \label{g12drange}
\end{eqnarray}
is fulfilled in any model of new physics which leaves $\Gamma_{12}^d$
unaffected.  Can new physics enhance $\dg_d$ to an observable level?
$\Gamma_{12}^d$ is an inclusive quantity with the three doubly
Cabibbo-suppressed contributions $\Gamma_{12}^{d,cc}$,
$\Gamma_{12}^{d,uc}$ and $\Gamma_{12}^{d,uu}$. The first contribution
stems from the interference of the decay $b\to c\ov c d$ of the
$\Bbar_d$ component with the decay $\ov b\to \ov c c \ov d$ of the $B_d$
component of the mixed neutral meson state. This interference is
possible, because both components can decay into the same flavourless
$c\ov c d\ov d$ final state. $\Gamma_{12}^{d,uu}$ is the analogue
involving $b\to u \ov u d$, while $\Gamma_{12}^{d,uc}$ arises from the
interference of $b\to c\ov u d$ and $\ov b\to \ov u c \ov d$ decays or
their charge-conjugate modes.  It is difficult to find a model of new
physics which can numerically compete with the dominant SM contribution
$\Gamma_{12}^{d,cc}$ without violating other experimental constraints.
For a recent discussion of the experimental aspects of $\dg_d$ see
Ref.~\cite{Gershon:2010wx}. Formulae for the time evolution of $B_d$
decays to flavourless states, which permit the extraction of $\dg_d$,
can be found in Refs.~\cite{ped,dfn,Nierste:2004uz}. {We update 
$\phi_d$ and $a_{\rm fs}^d$ below in \eq{phism}.}

In the $B_s$ system $\dg_s$ is found together with $\phi_s$ from an
angular analysis of $B_s \to J/\psi \phi$ data. The calculated value of
$|\Gamma_{12}^s|$ defines the physical ``yellow band'' in the $(\dg_s,
\phi_s)$ plane. In any model of new physics $(\dg_s, \phi_s)$ must lie
in this band, because new physics has a negligible impact on
$\Gamma_{12}^s$ which stems from Cabbibo-favoured $b\to c\ov c s$
decays. $|\Gamma_{12}^s|$ is proportional to the hadronic parameter 
$f_{B_s}^2 {\cal B}_{B_s}$. Updating our 2006 prediction in Ref.~\cite{ln} to 
the 2010 world averages of the input parameters listed in Tabs.~6 and 7
of Ref.~\cite{lnckmf} we find 
\begin{eqnarray}
\dg_s^{\rm SM} & \simeq & 2|\Gamma_{12}| \; =\;
\nn
&=&  
   \lt(\lt. {0.087} 
          \pm \lt.  {0.015}\rt|_{\widetilde{R}_2}
          \pm { 0.012}\rt|_{f_{B_s}} 
          \pm \lt.  {0.007}\rt|_{\rm scale} 
    \pm \lt.  {0.007}\rt|_{\rm rest} \rt)\, \mbox{ps}^{-1}
\label{ga12}
\end{eqnarray}
The three largest sources of uncertainty stem from the {matrix element
  of the operator} $\widetilde{R}_2$ which {occurs} at order $1/m_b$,
the decay constant $f_{B_s}$ and the choice of the renormalisation scale
(estimating higher-order corrections in $\alpha_s$).  Adding the errors
in quadrature yields
\begin{eqnarray}
\dg_s^{\rm SM} &\simeq & 2|\Gamma_{12}| \;=\;  
   ({0.087} \pm {0.021} )\, \mbox{ps}^{-1} .
\label{ga12qu}
\end{eqnarray}
The uncertainty of the theory prediction has reduced from 41\% in 2006
to {24\%} in 2010 because of an impressive progress in the lattice
calculations of $f_{B_s} \sqrt{{\cal B}_{B_s}}=212 \pm 14\,\mev$ (average from 
\cite{lnckmf} using \cite{latt1}). The corresponding
value used in Ref.~\cite{ln} was $f_{B_s} \sqrt{{\cal B}_{B_s}}=221 \pm 46\,\mev$.
The central value has decreased due to the smaller $f_{B_s} \sqrt{{\cal B}_{B_s}}$
and slightly smaller values of $m_b$, $V_{cb}$ and $\alpha_s$.
\eq{ga12qu} implies 
\begin{eqnarray}
\frac{\dg_s^{\rm SM}}{\Gamma_s}  &\simeq& \frac{2|\Gamma_{12}|}{\Gamma_s} 
 \;=\; {0.133 \pm 0.032} .
\label{degaga}
\end{eqnarray}
Here $\Gamma_s=\Gamma_d$ has been used, deviations from this relation
are discussed in the next section.  If one assumes that there is no new
physics in the mixing amplitude, one can determine $\dg_s^{\rm
  SM}/\Gamma_s$ in a more precise way:\footnote{{In Eq.~(3.27) of
    Ref.~\cite{ln} we have erroneously used $\tau_{B_s}^{\rm exp}$
    instead of $\tau_{B_d}^{\rm exp}$ in the calculation. The quoted
    number should read $\dg_s\tau_{B_d}=0.135\pm0.026$.}}
\begin{eqnarray}
\frac{\dg_s^{\rm SM}}{\Gamma_s} & = & 
\frac{\Delta \Gamma_s^{\rm SM}}{\Delta M_s^{\rm SM}} \cdot \Delta M_s^{\rm Exp.} 
\cdot \tau_{B_d}^{\rm Exp}
 \;=\; 0.137 \pm 0.027 .
\end{eqnarray}
The CP phases read 
\begin{equation}
\label{phism}
\begin{array}{cclcccl}
 \phi_s^{\rm SM} & = & {0.22^\circ} \pm {0.06^\circ}, & \qquad \qquad & 
 \phi_d^{\rm SM} & = & {-4.3^\circ} \pm 1.4^\circ. 
\end{array}
\end{equation}
The CP asymmetries in flavour-specific decays read
\begin{eqnarray}
 a_{\rm fs}^{s,\rm SM} = {  (1.9 \pm 0.3) \cdot 10^{-5}} ,\qquad &&\qquad 
 a_{\rm fs}^{d,\rm SM} = {- (4.1 \pm 0.6) \cdot 10^{-4}  } . \label{afssm}
\end{eqnarray}
{In \eqsand{phism}{afssm} the result
  $|V_{ub}|=(3.56\epm{0.15}{0.20})\cdot 10^{-3}$ of the SM fit in
  Ref.~\cite{lnckmf} has been used. If one uses instead the experimental
  value $|V_{ub}|=(3.92\pm 0.46) \cdot 10^{-3}$, one finds slightly
  larger values, e.g.\ $ a_{\rm fs}^{s,\rm SM}= (2.1 \pm 0.4) \cdot
  10^{-5} $ and $a_{\rm fs}^{d,\rm SM}= - (4.5 \pm 0.8) \cdot 10^{-4}$.}
{Experiments address different linear combinations of these two CP
  asymmetries:} The D\O\ collaboration has measured the dimuon asymmetry
$A_{SL}$ defined in Eq. (\ref{DEFdimuon}), while LHCb will measure the
difference of these asymmetries.  The corresponding Standard-Model
predictions read
\begin{eqnarray}
A_{SL}^{\rm SM} & = & -\left( {2.0 \pm 0.3}\right) \cdot 10^{-4} \; ,
\label{DimuonSM}
\\
a_{\rm fs}^{s,\rm SM} -  a_{\rm fs}^{d,\rm SM} & = &  
  \; \; \; { \left( 4.3 \pm 0.7\right)} \cdot 10^{-4} \; .
\label{AslDifferenceSM}
\end{eqnarray}
\\
{The central values in Eqs.~(\ref{dgdmd})-(\ref{AslDifferenceSM})
  correspond to a renormalisation scheme using $\ov{\rm MS}$ quark
  masses and $\ov z=\ov m_c^2(\ov m_b)/\ov m_b^2(\ov m_b)$. }  For
completeness we also update the NLO prediction \cite{bjw} for
$\dm_s^{\rm SM}$:
\begin{eqnarray}
\dm_s^{\rm SM} &=& (17.3 \pm {2.6}) \, \mbox{ps}^{-1} . 
\end{eqnarray}
The quoted central value corresponds to $f_{B_s}=231\,\mev$ and
  {${\cal B}_{B_s}=0.841$} (see Ref.~\cite{lnckmf}). Often $\dm_s^{\rm
    SM}$ is expressed in terms of the scheme-independent parameter
  {$\widehat{\cal B}_{B_s}$ and ${\cal B}_{B_s}=0.841$} translates to
  {$\widehat{\cal B}_{B_s}=1.281$.}

We finally discuss the width difference between the CP eigenstates 
defined as 
\begin{eqnarray}
\ket{B_{s,\rm CP\pm}} &=&
  \frac{\ket{B_s}\mp\ket{\Bbar_s}}{\sqrt2} . \no
\end{eqnarray}
The width of the CP-even state exceeds that of the CP-odd state by 
\begin{eqnarray}
\dg_{\rm CP} &=& 2 |\Gamma_{12}^s| 
\end{eqnarray}
and is unaffected by new physics in $M_{12}^s$ \cite{dfn}. Aleksan et
al.\ have shown that in the simultaneous limits $m_c\to \infty$, $m_b-2
m_c\to 0$ and an infinite number of colours, $N_c \to \infty$,
$\dg_{\rm CP}$ is exhausted by decays into just four final states 
\cite{Aleksan:1993qp}:
\begin{eqnarray}
  2\, B(\BsorBsbar \to D_s^{(*)}{}^+ D_s^{(*)}{}^-)
= \frac{\dg_{\rm CP}}{\Gamma_s} \lt[
  1 + {\cal O} \lt( \frac{\dg}{\Gamma_s} \rt) \rt] .\label{aleksan}
\end{eqnarray}
On the experimental side we have the Belle measurement \cite{Esen:2010jq} 
\begin{eqnarray}
\frac{\dg_{\rm CP}}{\Gamma_s} =
  \lt.\lt. 0.147 \epm{0.036}{0.030}\rt|_{\rm stat}
       \epm{0.044}{0.042}\rt|_{\rm syst} \no 
\end{eqnarray}
and the D\O\ result \cite{d0note} 
\begin{eqnarray}
\frac{\dg_{\rm CP}}{\Gamma_s} = 0.072 \pm 0.030 . \no
\end{eqnarray}
While it is flattering that the Belle central value exactly conicides
with our 2006 prediction for $\dg_{\rm CP}/\Gamma_s$ in Ref.~\cite{ln},
the data are not accurate enough to assess the accuracy of the limits
$m_c\to \infty$, $m_b-2 m_c\to 0$ and $N_c\to \infty$ adopted in
Ref.~\cite{Aleksan:1993qp}. However, the calculated $1/m_b$ corrections
are of order 20\% \cite{bbd,ln} and in the NLO result of
Refs.~\cite{bbgln,rome,ln} the $1/N_c$ terms are non-negligible. Also
the deviation of the hadronic ``bag'' factor $B\simeq 0.85$ from 1 is an
$1/N_c$ effect. Therefore one cannot rule out large corrections to
\eq{aleksan}, possibly of order 100\%. On the experimental side one may
look for multi-body $c\ov c s\ov s$ final states which are absent in the
limit $N_c\to \infty$. One ingredient of \eq{aleksan} is the prediction
that $B_{s,\rm CP-}$ does not contribute to $\BsorBsbar \to
D_s^{(*)}{}^+ D_s^{(*)}{}^-$. This can be checked by studying the
lifetime in these modes, which should then be equal to $1/\Gamma_L^s$ as
measured in the CP-even component of $\BsorBsbar \to J/\psi \phi$
\cite{dfn}.

\boldmath 
\section{Average $B_s$ width}\label{sec:gs}
\unboldmath 
We define 
\begin{eqnarray}
 \tau_{B_s} \equiv \frac{1}{\Gamma_{B_s}}.  
\end{eqnarray}
Any decay $\BsorBsbar \to f$ obeys a two-exponential law,
$\Gamma[\BsorBsbar \to f,t] \to A_f \exp[-\Gamma_L^s t] + B_f
\exp[-\Gamma_H^s t]$. In a flavour-specific decay like $B_s \to X\ell^+
\nu_\ell$ or $B_s \to D_s^- \pi^+$ the two coefficients are equal,
$A_f=B_f$. Fitting the decay to a single exponential, one determines
$\tau_{B_s}$ up to a calculable correction of order $\dg^2/\Gamma_s^2$
\cite{hm,dfn}. If the experimental selection efficiencies vary over the
decay length, this method can lead to a bias towards $\Gamma_L^s$ or
$\Gamma_H^s$, therefore it is recommended to use the correct
two-exponential formulae in the experimental analyses with simultaneous
fits to $\Gamma_s$ and $\dg_s$. 

\begin{figure}[tb]
\centering
\includegraphics[height=0.15\textwidth]{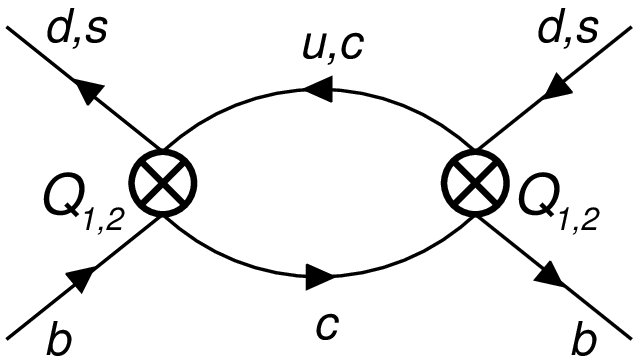} \hspace{3cm}
\includegraphics[height=0.15\textwidth]{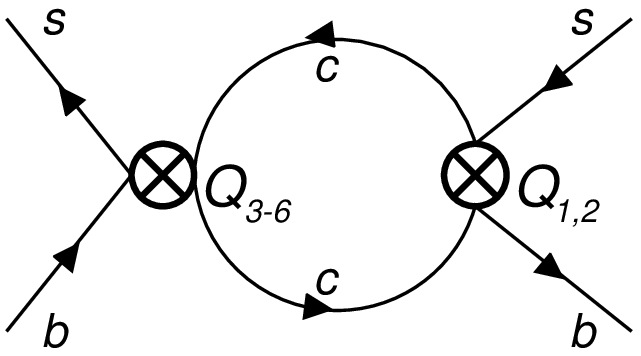}\\[3mm]
\includegraphics[height=0.15\textwidth]{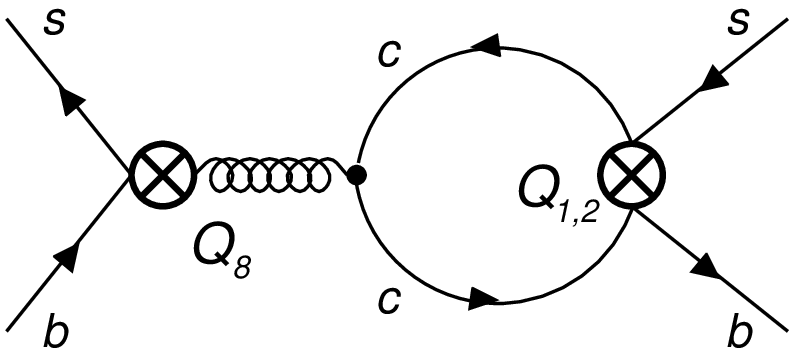}\hspace{2cm}
\includegraphics[height=0.15\textwidth]{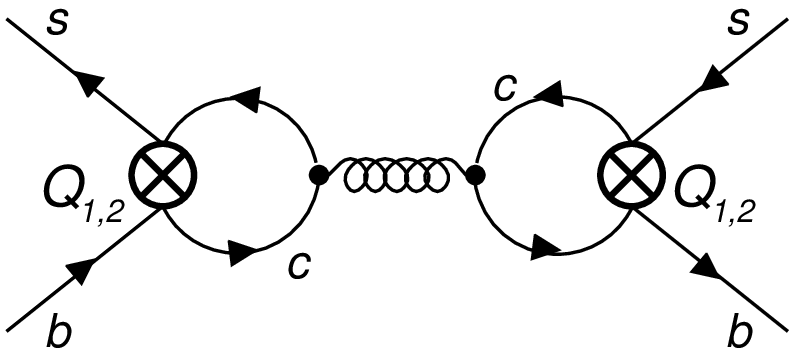}
\caption{\it CKM-favoured weak annihilation diagrams: The upper left diagram
  contributes nearly equally to $\tau_{B_s}$ and $\tau_{B_d}$, the other
  diagrams involve small penguin coefficients $C_{3-6}$ or are
  suppressed by $\alpha_s$. }
\label{fig:wa}
\end{figure}

We next discuss the ratio $\tau_{B_s}/\tau_{{B_d}}=\Gamma_d/\Gamma_s$: The
deviation of $\tau_{B_s}/\tau_{{B_d}}$ from 1 stems from the weak
annihilation (WA) diagrams of \fig{fig:wa}.
The upper left diagram \cite{spec,ns,bbd} involves the large Wilson
coefficients $C_{1,2}$, but this contribution almost cancels from
$\tau_{B_s}/\tau_{B_d}$ up to terms of order $z=m_c^2/m_b^2$ and
$1-f_{B_d}^2M_{B_d}/f_{B_s}^2M_{B_s}$. The other diagrams
\cite{goldesel} essentially only contribute to $\tau_{B_s}$, but involve
small coefficients or a factor of $\alpha_s(m_b)$. Since moreover WA
diagrams are individually small, $|\tau_{B_s}/\tau_{B_d}-1|$ has been
estimated to be of order 0.01 or smaller by several authors
\cite{spec,ns,bbd,goldesel}.  (There is also a contribution to
$\tau_{B_s}/\tau_{B_d}$ from SU(3)$_{\rm F}$ violation in the
kinetic-energy and chromomagnetic operators, which is of order $10^{-3}$
and negligible \cite{bbd}.) The theory prediction involves four hadronic
parameters \cite{ns}, $B_1$, $B_2$, $\epsilon_1$ and $\epsilon_2$, which
are multiplied by the $B_s$ meson decay constant $f_{B_s}$, and further
the ratio $f_{B_s}/f_{B_d} = 1.209 \pm 0.007 \pm 0.023$
\cite{lnckmf,latt2}.

We find 
\begin{eqnarray}
 \frac{\tau_{B_s}}{\tau_{B_d}}-1 &=&   10^{-3} \cdot
\lt( \frac{f_{B_s}}{231\,\mev}\rt) ^2 \; \big[ \,
  (0.77\pm 0.10) B_1 - (1.00\pm 0.13)  B_2  \nn
&&  \qquad\qquad \qquad \qquad 
        + (36\pm 5)  \epsilon_1 - (51 \pm 7) \epsilon_2 \, \big] .
\end{eqnarray}
Using the results of the quenched lattice-QCD calculation for $B_{1,2}$
and $\epsilon_{1,2}$ of Ref.~\cite{Becirevic:2001fy}, $(B_1,B_2,
\epsilon_1,\epsilon_2) = (1.10\pm 0.20, \, 0.79 \pm 0.10, \, -0.02 \pm
0.02, \, 0.03 \pm 0.01 )$, we find {$-4\cdot 10^{-3} \leq
  \frac{\tau_{B_s}}{\tau_{B_d}}-1 \leq 0$.} The lower end of this
interval exceeds the HFAG value of $\tau_{B_s}/\tau_{B_d}-1 \, =\,  −- 0.027
\pm 0.015$ \cite{hfag} by 1.5 standard deviations.

\begin{figure}
\centering
\includegraphics[width=0.8\textwidth, angle = 0]{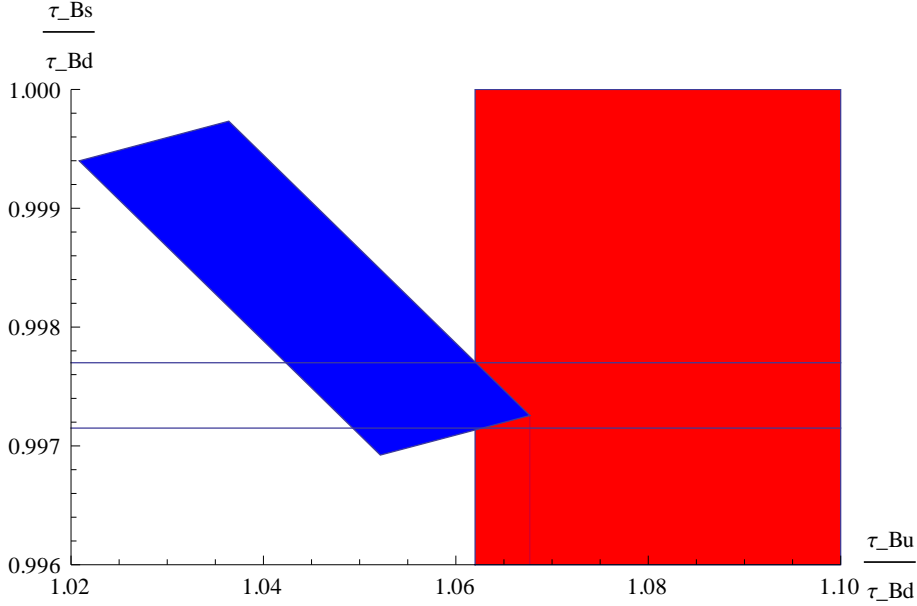}
\caption{\it Allowed range for $\tau_{B_s}/\tau_{B_d}$ and
  $\tau_{B^+}/\tau_{B_d} $. {The predictions for $\tau_{B^+}/\tau_{B_d} $
    and $\tau_{B_s}/\tau_{B_d}$ (blue region)
    derived with the decade-old hadronic parameters of
    Ref.~\cite{Becirevic:2001fy} barely overlaps with the experimental
    3$\sigma$ region {(red region to the right)}. This shows the
    importance of a modern lattice calculation of these
    parameters.}}\label{fig:corr}
\end{figure}

We can improve our prediction by using the information of the lifetime
difference between $B^+$ and $B_d$ meson, which involves the same
hadronic parameters, up to SU(3)$_{\rm F}$ corrections.  Adding $1/m_b$
corrections to the prediction in Ref.~\cite{bbgln2} and using the input
parameters of Ref.~\cite{lnckmf} we obtain
\begin{eqnarray} 
  \frac{\tau_{B^+}}{\tau_{B_d}} - 1 
&=& {0.0324}   
           \lt( \frac{f_B}{200\mev} \rt)^2 \,
      \Big[ \, ( 1.0 \pm 0.2) \, B_1 \; + \; (0.1 \pm 0.1) \, B_2 \nn 
 &&     \phantom{  %0.0324
           \lt( \frac{f_B}{200\mev} \rt)^2 \,
      \Big[} \; - \;
         {(17.8 \pm 0.9)} \, \epsilon_1 \; + \; (3.9 \pm 0.2) \, \epsilon_2
      - 0.26 \, \Big] .~~~ \label{phent}
\end{eqnarray}
The last term are the $1/m_b$ corrections calculated in the vacuum
  saturation approximation.
With the hadronic parameters of Ref.~\cite{Becirevic:2001fy} quoted above we
correlate $\tau_{B_s}/\tau_{B_d}$ with $\tau_{B^+}/\tau_{B_d} $ in
Fig.~\ref{fig:corr}.
We notice that the experimental range $\tau_{B^+}/\tau_{B_d}=1.081 \pm
0.006$ \cite{hfag} prefers {a larger deviation} of 
$\tau_{B_s}/\tau_{B_d}$ {from 1.} Our analysis omits effects of
SU(3)$_{\rm F}$ violation other than
$1-f_{B_d}^2M_{B_d}/f_{B_s}^2M_{B_s}$; SU(3)$_{\rm F}$ violation in
$B_1,B_2,\epsilon_1,\epsilon_2$ may lower
$\tau_{B_s}/\tau_{B_d}$ {a bit further} 
\cite{Becirevic:2008us}.  {Most importantly, we observe a tension
  between theory and experiment in $\tau_{B^+}/\tau_{B_d}$. This calls for
  a new effort in the calculation of $B_1,B_2,\epsilon_1,$ and
  $\epsilon_2$ on the lattice. We further note that with the 2001 values
  of these hadronic parameters the theory prediction for
  $\tau_{B_s}/\tau_{B_d}$ is in better agreement with experiment
  than that of $\tau_{B^+}/\tau_{B_d}$.}

\boldmath 
\section{New physics in \bbm}\label{sec:np}
\unboldmath 
New physics in \bbm\ can be parametrised in terms of the complex
parameters $\Delta_d$ and $\Delta_s$ defined as  \cite{ln}
\begin{eqnarray}
\Delta_q &\equiv & \frac{M_{12}^q}{M_{12}^{\text{SM},q}},
\qquad\qquad  \Delta_q \; \equiv \;  |\Delta_q| e^{i \phi^\Delta_q} .
 \label{defdel}
\end{eqnarray}
% The CP phase of \eq{defphi} reads
% \begin{eqnarray}
% \phi_q &=& \phi_q^{\text{SM}} + \phi^\Delta_q . \label{phiq}
% \end{eqnarray}
% and the SM contributions in \eq{phism} are negligible in view of
% current uncertainties. 
The average of the D\O\ \cite{dimuon_evidence_d0} and CDF 
\cite{ASLCDF} measurements of $A_\text{SL}$, 
\begin{equation}
       A_\text{SL}=-0.0085 \pm 0.0028, \label{aslavg}
\end{equation}
is 2.9$\sigma$ away from the SM prediction. Since $ A_\text{SL}$
involves both \bbms\ and \bbmd, a combined analysis of $\Delta_d$ and
$\Delta_s$ is desirable. For the SM prediction of the two key
observables related to \bbmd, $\dm_d$ and $A_\text{CP}^\text{mix}
(B_d\to J/\psi K_S)$, one needs the information on the apex
$(\ov{\rho},\ov{\eta})$ of the unitarity triangle (UT).  $\ov{\rho}$ and
$\ov{\eta}$ in turn depend on $\Delta_d$ and, through $\dm_d/\dm_s$,
even on $|\Delta_s|$. These interdependences require a joint fit to the
CKM elements and $\Delta_d$ and $\Delta_s$. Further, theoretical
assumptions on the nature of new physics are needed: A plausible
framework is the hypothesis that observable effects of new physics are
confined to \bb\ and \kkm, which are typically more sensitive to new
physics than the other quantities entering the global fit of the
UT. 

\begin{figure}
\includegraphics[width=0.48\textwidth]{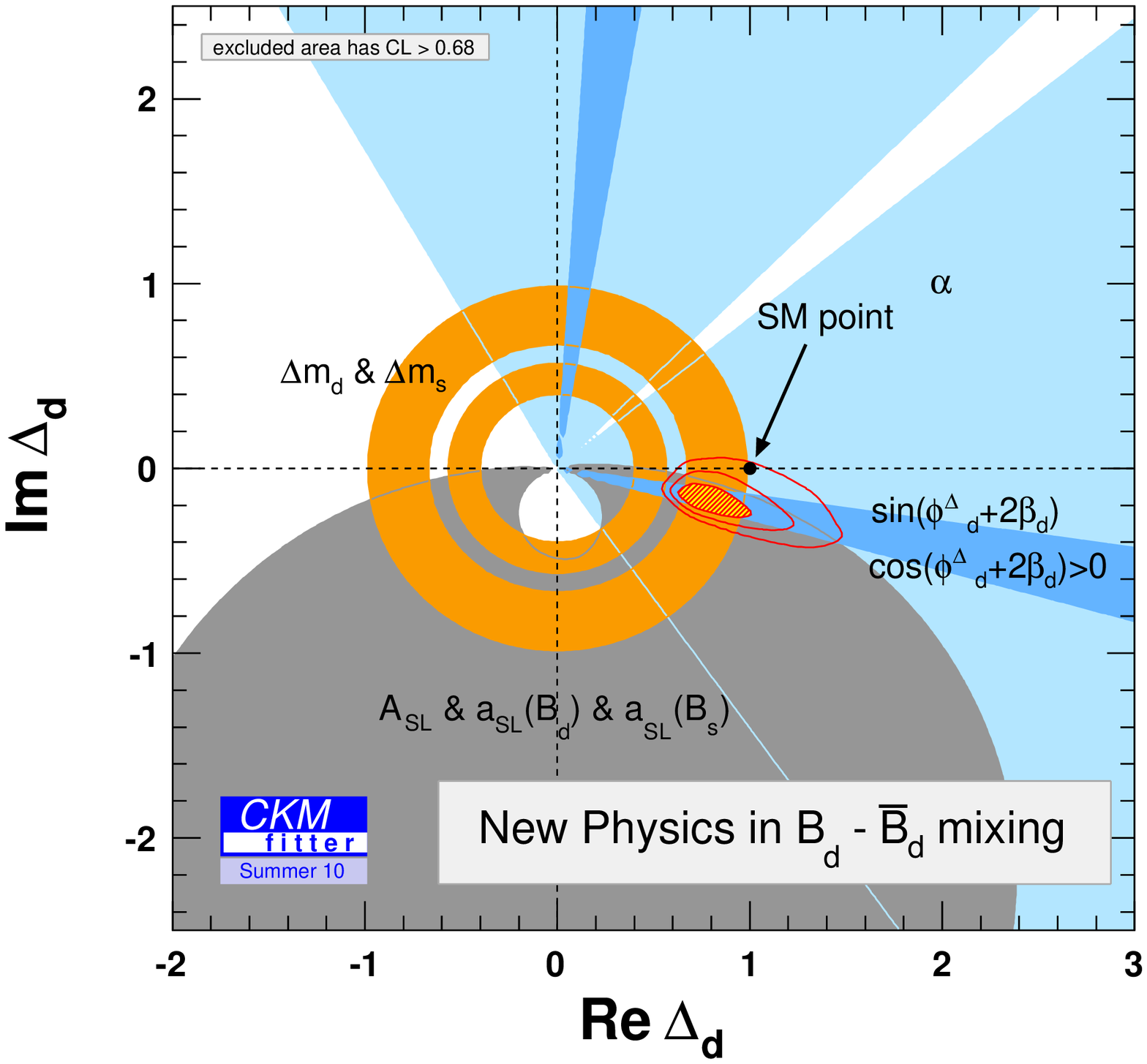}
\hfill
\includegraphics[width=0.48\textwidth]{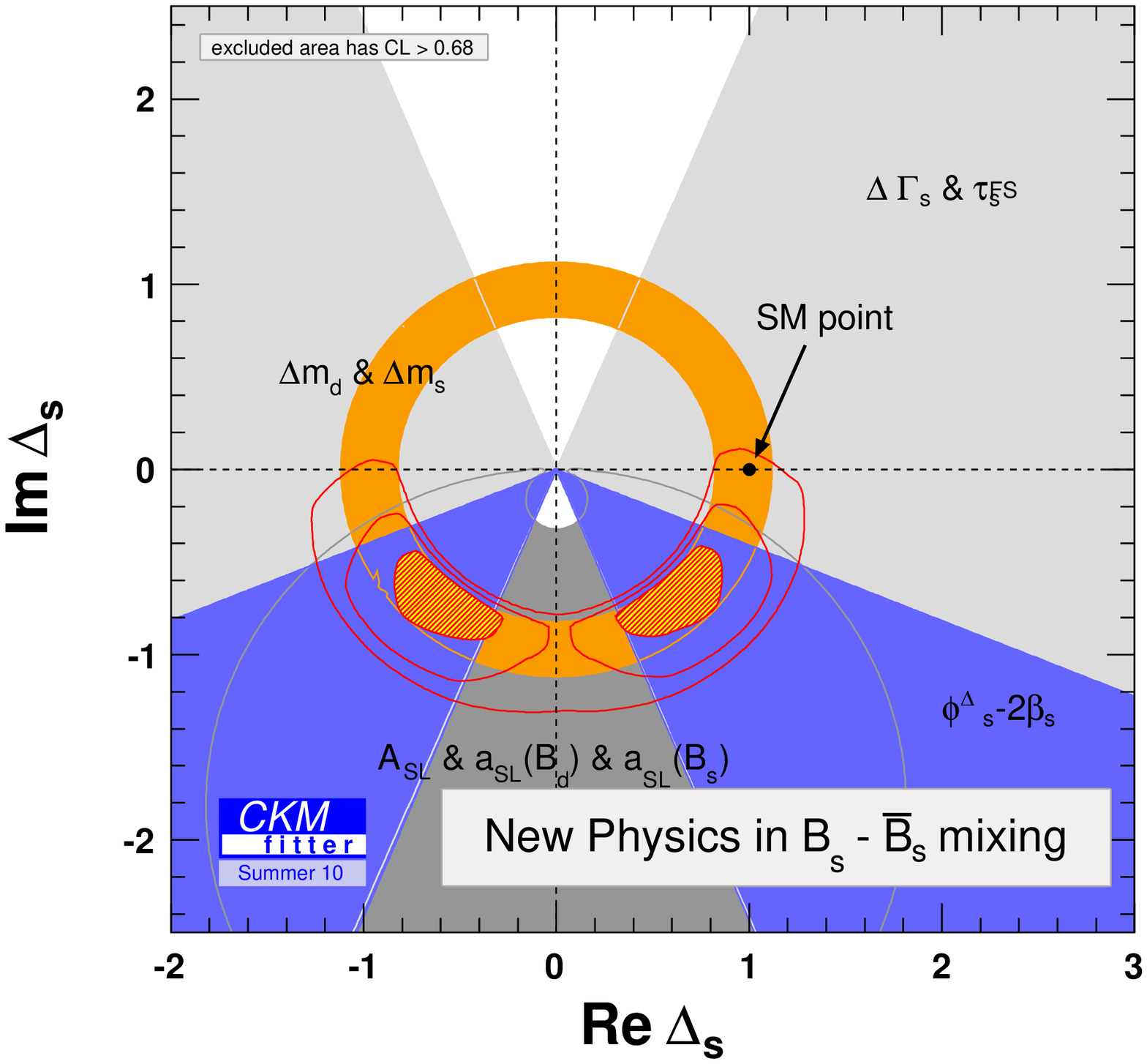}
  \caption{\it Allowed regions for $\Delta_d$ and $\Delta_s$ \cite{lnckmf}.}\label{fig:res}
\end{figure}

Here we briefly report on our analysis with the CKMfitter
collaboration, restricting ourselves to the first of three studied
scenarios, and refer to Ref.~\cite{lnckmf} for details. This scenario 
treats  $\Delta_d$ and $\Delta_s$ as independent quantities. 
The fit results are shown in Fig.~\ref{fig:res}.
The tension on the SM seen in $\Delta_d\neq 1$ is driven by several
BaBar and Belle measurements \cite{babe} of $B(B^+\to \tau^+ \nu_\tau)$,
which is proportional to $|V_{ub}|^2 f_B^2$. The different
determinations of $|V_{ub}|\propto \sqrt{\ov{\rho}^2+\ov{\eta}^2}$ are
consistent with each other, but the corresponding best-fit region in the
$(\ov{\rho},\ov{\eta})$ plane is not spiked by the ray stemming from
$\beta\simeq 21^\circ$ measured through $A_\text{CP}^\text{mix} (B_d\to
J/\psi K_S)$. In the presence of new physics in \bbmd\ the latter
quantity determines $\sin (2\beta+\phi_d)$ and $\phi_d<0$ (favoured in
Fig.~\ref{fig:res}) alleviates the tension. Negative values of $\phi_d$
further help to accomodate the large negative result for $A_\text{SL}$.
Not only $A_\text{SL}$ but also both the D\O\ and CDF measurements of
$\phi_s^\Delta$ through $B_s\to J/\psi \phi$ favour
$\phi_s^\Delta<0$. Our best-fit value is
$\phi_s^\Delta=(-52\epm{32}{25})^\circ$ at $ 95\%\,$CL. This is
consistent with the 2010 results found from $B_s\to J/\psi \phi$,
$\phi_s^{\Delta,CDF}=(-29 \epm{44}{49})^\circ$ and
$\phi_s^{\Delta,D\O}=(-44 \epm{59}{51})^\circ$ at $ 95\%\,$CL each. If
we remove $A_\text{SL}$ from the fit and predict it instead from the
other observables through the global fit, we find $A_{\rm SL}=\lt( -4.2
\epm{2.9}{2.7}\rt)\cdot 10^{-3}$ at $ 95\%\,$CL, which is just
1.5$\,\sigma$ away from the experimental number in \eq{aslavg}. The
new-physics scenario with $\Delta_{d,s} \neq 1$ gives an excellent
fit, with the SM point $\Delta_{d}=\Delta_{s}=1$ excluded at the level
of 3.6 standard deviations {\cite{lnckmf}}. We remark that this result is
only marginally influenced by uncertainties of lattice calculations.
{Using the fit results of \cite{lnckmf} for scenario I 
we can {also quote a} number for the quantity
\begin{eqnarray}
S_{J/\psi \phi} & = & \sin \left(-2\beta_s + \Phi_s^\Delta \right)
               \; = \; -0.78^{+0.19}_{-0.12}\; 
\end{eqnarray}
{used in theoretical papers. We have assumed the absence of new
  physics in the tree-dominated decay $B_s\to J/\psi \phi$.}
This value differs sizeably from the Standard-Model expectation of
\begin{eqnarray}
S_{J/\psi \phi}^{\rm SM} & = & -0.036 \pm 0.002 \; .
\end{eqnarray}
In the literature {numerous extensions of the Standard Model are
  discussed} to explain this difference, see e.g. \cite{lnckmf,NPfromus}
and references therein.}

In conclusion we have updated several theory predictions related to 
$B$ meson lifetimes and \bbm\ quantities. We have further briefly
discussed the essential results of our study with the CKMFitter group 
\cite{lnckmf}, which has found evidence of physics beyond the SM in 
\bbm.

%%%%%%%%%%%%%%%%%%%%%%%%%%%%%%%%%%%%%%%%%%%%%%%%%%%%%%%%%%%%%%%%%%%%%%%%%
%%
%%   use this format to include an .eps figure into your paper
%%

\Acknowledgements
UN thanks the conference organisers for their invitation and financial 
support. The presented work is supported by BMBF grant 05H09VKF.

\section*{Appendix: Theory errors }\label{app}
In this appendix we give a detailed list of the different sources of the
theoretical error for observables in the $B_s$ mixing system.  We
compare this numbers with the corresponding ones from Ref.~\cite{ln},
{but slightly proceed in a different way: In accordance with
  Ref.~\cite{lnckmf} we use the $\ov{\rm MS}$ scheme for $m_b$, while
  {in \cite{ln} we were using in addition the pole scheme and our
    numbers and errors were averages of these two quark mass schemes.}
  The numerical values and uncertainties of the input parameters are
  taken from Table 6 and Table 7 of Reference \cite{lnckmf}. Contrary to
  Ref.~\cite{lnckmf} we do not use the Rfit method for the statistical
  analysis in these proceedings, instead we simply add different
  uncertainties in quadrature.}

{$\ov m_c(\ov m_c)=(1.286\pm 0.042) \,\gev$ and $\ov m_b(\ov m_b)=(4.248 \pm
  0.051)\,\gev$ imply }  
\begin{equation}
\ov z = \frac{\ov m_c^2(\ov m_b)}{\ov m_b^2(\ov m_b)} = 0.0474 \pm 0.0033 \; .
\end{equation}
For the CKM angle $\gamma$ we do not use the direct measurement given in Table
6 of Ref.~\cite{lnckmf}, but instead the direct bounds on
$\alpha^{\rm exp}$ and $\beta^{\rm exp}$ from the same table:
\begin{eqnarray}
\gamma & = & \pi - \alpha^{\rm exp} - \beta^{\rm exp}
       \; = \; 1.220 \pm 0.077 \; =\; 69.9^\circ \pm 4.4^\circ   
     . \label{gamval}
\end{eqnarray}
{In this way we find a precise value for $\gamma$ from which new
  physics in \bbm\ drops out. New physics can only change the value in
  \eq{gamval} through novel electroweak $b\to d$ penguin effects, which
  stay within the quoted error in all plausible models. We calculate the
  CKM elements using $V_{us}$, $V_{cb}$, $|V_{ub}/V_{cb}|$ and $\gamma$
  as inputs.}

{The error budget for $\dm_s^{\rm SM}$ is as follows:}
\begin{displaymath}
\begin{array}{|c||c|c|}
\hline
\Delta M_s^{\rm SM}            &  \mbox{this work}       &  {\mbox{Ref.~\cite{ln}}}
\\
\hline
\hline
\mbox{Central Value}   &  17.3 \, \mbox{ps}^{-1} &  19.3 \, \mbox{ps}^{-1}
\\
\hline
\delta (f_{B_s})       &  13.2 \%                &  33.4 \% 
\\
\hline
\delta (V_{cb})        &    3.4 \%               &  4.9 \%
\\
\hline
\delta ({\cal B}_{B_s}) &   2.9 \%                &  7.1 \%
\\
\hline
\delta (m_t)           &   1.1 \%                &  1.8 \%
\\
\hline
\delta (\alpha_s)     &   0.4 \%                 &  2.0 \%
\\
\hline
\delta (\gamma)        &   0.3 \%                &  1.0 \%
\\
\hline
\delta (|V_{ub}/V_{cb}|) &   0.2 \%                &  0.5 \%
\\
\hline
\delta (\ov m_b)           &   0.1 \%                &  ---
\\
\hline
\hline
\sum \delta            &  14.0 \%                & 34.6 \%
\\
\hline
\end{array}
\end{displaymath}
For the mass difference we observe a considerable reduction of the
overall error from {$35\%$} in 2006 to $14 \%$ {today}. This is
mainly driven by the progress in the lattice determination of 
{$f_{B_s}^2 {\cal B}_{B_s}$}. {The situation for $\dg_s^{\rm SM}$ is similar:}
\begin{displaymath}
\begin{array}{|c||c|c|}
\hline
\Delta \Gamma_s^{\rm SM} &  \mbox{this work}  & {\mbox{Ref.~\cite{ln}}}  
\\
\hline
\hline
\mbox{Central Value}   &  0.087 \, \mbox{ps}^{-1} &  0.096 \, \mbox{ps}^{-1}
\\
\hline
\delta ({\cal B}_{\widetilde R_2})       &   17.2 \%                &  15.7 \%
\\
\hline
\delta (f_{B_s})       &  13.2 \%                 &  33.4 \% 
\\
\hline
\delta (\mu)           &   7.8 \%                 &  13.7 \%
\\
\hline
\delta (\widetilde{{\cal B}}_{S,B_s})           &   4.8 \%                 &  3.1 \%
\\
\hline
\delta ({\cal B}_{R_0})       &    3.4 \%                &   3.0 \%
\\
\hline
\delta (V_{cb})        &    3.4 \%                &  4.9 \%
\\
\hline
\delta ({\cal B}_{B_s})             &   2.7 \%                 &  6.6 \%
\\
\hline
\delta ({\cal B}_{\tilde{R}_1}) &  1.9 \%               &   ---
\\
\hline
\delta (\bar z )       &    1.5 \%               &   1.9 \%
\\
\hline
\delta (m_s )          &    1.0 \%               &   1.0 \%
\\
\hline
\delta ({\cal B}_{R_1})       &    0.8 \%               &   ---
\\
\hline
\delta ({\cal B}_{\tilde{R}_3}) &  0.5 \%               &   ----
\\
\hline
\delta (\alpha_s)      &   0.4 \%                 &  0.1 \%
\\
\hline
\delta (\gamma)        &   0.3 \%                &  1.0 \%
\\
\hline
\delta ({\cal B}_{R_3})       &   0.2 \%               &  ---
\\
\hline
\delta (|V_{ub}/V_{cb}|) &   0.2 \%                &  0.5 \%
\\
\hline
\delta (m_b )          &   0.1 \%               &   1.0 \%
\\
\hline
\hline
\sum \delta            &  24.5 \%                & 40.5 \%
\\
\hline
\end{array}
\end{displaymath}
For the decay rate difference we also find a strong reduction of the
overall error from $40.5\%$ in 2006 to $24.5 \%$. This is again due to
our more precise knowledge of the decay constant and the bag parameter
${\cal B}_{B_s}$.  {However, in the $\overline{\rm MS}$-scheme for the
  quark masses also the renormalisation-scale dependence is reduced.}
% In \cite{ln} we were using in addition the pole scheme, and our numbers
% and errors were averages of these two quark mass schemes.  
It is interesting to note that now the dominant uncertainty stems
from the value of the matrix element of the power-suppressed operator
$\tilde R_2$ parametrised by ${\cal B}_{\tilde R_2}$. 

{Next we discuss the ratio of $\dg_s^{\rm SM} / \dm_s^{\rm SM}$:}
\begin{displaymath}
\begin{array}{|c||c|c|}
\hline
\Delta  \Gamma_s^{\rm SM} / \Delta M_s^{\rm SM}  &  \mbox{this work}       
        & {\mbox{Ref.~\cite{ln}}}
\\
\hline
\hline
\mbox{Central Value}   &  50.4 \cdot 10^{-4}        &  49.7 \cdot 10^{-4} 
\\
\hline
\delta ({\cal B}_{R_2})       &   17.2 \%                &  15.7 \%
\\
\hline
\delta (\mu)           &   7.8 \%                 &   9.1 \%
\\
\hline
\delta (\widetilde{{\cal B}}_{S,B_s})           &   4.8 \%                 &  3.1 \%
\\
\hline
\delta ({\cal B}_{R_0})       &    3.4 \%                &   3.0 \%
\\
\hline
\delta ({\cal B}_{\tilde{R}_1}) &  1.9 \%               &   ---
\\
\hline
\delta (\bar z)        &   1.5 \%                 &   1.9 \%
\\
\hline
\delta (m_b)           &   1.4 \%                 &   1.0 \%
\\
\hline
\delta (m_t)           &   1.1 \%                 &   1.8 \%
\\
\hline
\delta (m_s)           &   1.0 \%                 &   0.1 \%
\\
\hline
\delta (\alpha_s)      &   0.8 \%                 &  0.1 \%
\\
\hline
\delta ({\cal B}_{R_1})       &    0.8 \%                &   ---
\\
\hline
\delta ({\cal B}_{\tilde{R}_3}) &  0.5 \%                &   ----
\\
\hline
\delta ({\cal B}_{R_3})       &   0.2 \%                 &  ---
\\
\hline
\delta ({\cal B}_{B_s})             &   0.1 \%                 &  0.5 \%
\\
\hline
\delta (\gamma)        &   0.0 \%                &  0.1 \%
\\
\hline
\delta (|V_{ub}/V_{cb}|) &   0.0 \%                &  0.1 \%
\\
\hline
\delta (V_{cb})        &    0.0 \%                & 0.0 \%
\\
\hline
\hline
\sum \delta            &  20.1 \%                & 18.9 \%
\\
\hline
\end{array}
\end{displaymath}
For $\dg_s / \dm_s$ we do not find any improvement.  The decay constant
cancels out in this ratio and therefore we do not profit from the
progress in lattice simulations. Also the CKM dependence cancels to a
large extent.
% The improvement in the renormalisation scale dependence is less pronounced than 
% in $\Delta \Gamma$ alone.
{Our last error budget concerns the CP asymmetry in flavour-specific
$B_s$ decays:}
\begin{displaymath}
\begin{array}{|c||c|c|}
\hline
a_{\rm fs}^{s,{\rm SM}}               &  \mbox{This work}       &  {\mbox{Ref.~\cite{ln}}}
\\
\hline
\hline
\mbox{Central Value}   &  2.11 \cdot 10^{-5}        &  2.06 \cdot 10^{-5} 
\\
\hline
\delta (|V_{ub}/V_{cb}|) &  11.6 \%                   &  19.5 \%
\\
\hline
\delta (\mu)           &   8.9 \%                   &  12.7 \%
\\
\hline
\delta (\bar z)        &   7.9 \%                   &  9.3 \%
\\
\hline
\delta (\gamma)        &   3.1 \%                   &  11.3 \%
\\
\hline
\delta ({\cal B}_{\tilde{R}_3}) &  2.8 \%                  &   2.5 \%
\\
\hline
\delta (m_s)           &   2.0 \%                   &  3.7 \%
\\
\hline
\delta (\alpha_s)      &   1.8 \%                   &  0.7 \%
\\
\hline
\delta ({\cal B}_{R_3})       &   1.2 \%                   &  1.1 \%
\\
\hline
\delta (m_t)           &   1.1 \%                   &  1.8 \%
\\
\hline
\delta (\widetilde{{\cal B}}_{S,B_s})           &   0.6 \%                   &  0.4 \%
\\
\hline
\delta ({\cal B}_{R_0})       &    0.3 \%                  &   ---
\\
\hline
\delta ({\cal B}_{\tilde{R}_1}) &  0.2 \%                  &   ---
\\
\hline
\delta ({\cal B}_{B_s})             &   0.2 \%                   &  0.6 \%
\\
\hline
\delta (m_s)           &   0.1 \%                   &  0.1 \%
\\
\hline
\delta ({\cal B}_{R_2})       &   0.1 \%                   &  ---
\\
\hline
\delta ({\cal B}_{R_1})       &    0.0 \%                  &   ---
\\
\hline
\delta (V_{cb})        &    0.0 \%                  & 0.0 \%
\\
\hline
\hline
\sum \delta            &  17.3 \%                   & 27.9 \%
\\
\hline
\end{array}
\end{displaymath}
Finally we also observe a large improvement for the {CP asymmetry
  $a_{\rm fs}^s$}.  The overall error went down from $27.9\%$ to
$17.3\%$. In $a_{\rm fs}^s$ also the decay constant cancels, but in
contrast to $\dg_s/ \dm_s$ we now have a strong dependence on the CKM
elements. Here we benefit from more precise values of the CKM elements.
% and also from a more sizeable reduction of the renormalisation scale dependence. 
{The central value for $a_{\rm fs}^s$ and the error are different from
  the one quoted in \eq{afssm}, because in the table above we have
  computed the CKM elements from the experimental range for $|V_{ub}|$
  {centered around $|V_{ub}|=3.92\cdot 10^{-3}$ (see Tab.~6 of
    Ref.~\cite{lnckmf}).}  In \eq{afssm} the CKM elements are calculated
  instead from the more precise result
  $|V_{ub}|=(3.56\epm{0.15}{0.20})\cdot 10^{-3}$ of the global CKM fit.}

%%%%%%%%%%%%%%%%%%%%%%%%%%%%%%%%%%%%%%%%%%%%%%%%%%%%%%%%%%%%%%%%%%%%%%%%%%%%%%%%%%

\end{document}